\documentstyle[aps,multicol,prl,epsfig]{revtex}
\begin{document}
\def\prop{\propto}
\draft
\widetext
\title {Importance of matrix elements in the ARPES spectra of BISCO }
\author{A. Bansil$^1$ and M. Lindroos$^{1,2}$ }
\address{$^1$ Physics Department, Northeastern University,
Boston, Massachusetts 02115}
\address{$^2$ Tampere
University of Technology, P.O. Box 692, FIN-33101 Tampere, Finland}
\date{\today}
\maketitle
\widetext
\begin{abstract}

We have carried out extensive first-principles angle-resolved 
photointensity (ARPES) simulations in Bi2212 wherein the photoemission 
process is modelled realistically by taking into account the full crystal 
wavefunctions of the initial and final states in the presence of the 
surface. The spectral weight of the ARPES feature associated with the 
$CuO_2$ plane bands is found to undergo large and systematic variations 
with $k_\parallel$ as well as the energy and 
polarization of the incident photons. These theoretical predictions 
are in good accord with the corresponding measurements, 
indicating that the remarkable observed 
changes in the spectral weights in Bi2212 are essentially a matrix 
element effect and that the importance of matrix elements should be 
kept in mind in analyzing the ARPES spectra in the high-Tc's. 

\end{abstract}
\pacs{PACS numbers: 79.60.bm,71.18.+y,74.72.Hs}
\begin{multicols}{2}
\narrowtext

Angle-resolved photoemission spectroscopy (ARPES) has contributed 
significantly towards an understanding of the nature of the 
normal as well as the superconducting 
state of the cuprates\cite{gen,cape,stand,santa}. 
Much of the data on the cuprates, however, has been analyzed 
by assuming that the ARPES essentially measures the 
one-particle spectral function of the initial states. 
While this simple approach yields insights into the 
underlying physics, a satisfactory description of the 
spectra must necessarily model the photo-excitation process
properly by taking into account the matrix element involved, 
the complex modifications of the wavefunctions resulting from 
a specific surface termination, and the effects of multiple 
scattering and of finite lifetimes of the initial and final states. 

This article presents first-principles simulations of 
ARPES spectra in Bi2212 (BISCO) using the one-step model of 
photoemission which incorporates the aforementioned effects 
realistically\cite{Pendry2,mlcape,ncco,cucis,prl,stanf}. We 
focus on the ARPES signature of $CuO_2$ 
plane bands which are widely believed to be the key 
to the mechanism of superconductivity in the cuprates. 
The spectral weight of the ARPES peak
associated with the $CuO_2$ plane bands is found to undergo
large variations with ${\bf k}_\parallel$ as well as 
the energy and polarization of the incident photons. 
These theoretical predictions 
are in remarkable accord with the corresponding 
measurements\cite{foot1} on BISCO, and show clearly the importance of
"matrix element effects"\cite{foot2} in the ARPES spectra. Notably, 
a substantial increase in the ARPES spectral weight in BISCO 
in going from $\overline{\Gamma}$ to $\overline{M}$ was noted early by 
Anderson\cite{anderson1}, who speculated that this puzzling 
behavior may be the hallmark of spin charge separation. 

The physical and formal underpinnings of our approach 
may be exposed by starting with the following Golden rule-based 
expression for photointensity from initial states at energy 
E with photons of energy $\hbar\omega$:\cite{Schaich}
\begin{eqnarray}
I(E,\hbar\omega) = \frac{2\pi e}{\hbar}\sum\limits_{i,f}  
\vert < \Psi _f |\Delta|\Psi _i>|^2 \delta (E_f-E_i-\hbar\omega )
\end{eqnarray}
$\Psi_i (\Psi_f) $ are the initial (final) states 
of the semi-infinite solid, and 
$ \Delta = {e\hbar}/{2mc}({\bf p \cdot A + A \cdot p})$
is the interaction Hamiltonian with the 
electron momentum operator ${\bf p}$ and the 
vector potential ${\bf A}$ of the photon field. 

In the one-step model used in the present computations, Eq. 1 is 
manipulated into the form \cite{Pendry2}
\begin{eqnarray}
I({\bf k}_{\parallel},E,\hbar\omega)&=&
-\frac{1}{\pi} Im<{\bf k}_{\parallel}|G^+_2(E+\hbar\omega)\Delta 
 G^+_1(E)\Delta^{\dagger}
    \nonumber \\
    && {} \times G^-_2(E+\hbar\omega)|{\bf k}_{\parallel}>
	         \;
		 \label{H}
		 \end{eqnarray}
where the matrix element involves the free electron final state 
of momentum ${\bf k}_{\parallel}$. $G_2$ and $G_1$ denote the 
retarded (+) or advanced (-) one-electron Green
functions at appropriate energies. 

Notably, the so-called three-step model of photoemission approximates 
the matrix element in Eq. 1 via the {\it bulk} Bloch wavefunctions 
yielding for the photointensity {\it within} the solid\cite{Kunz}:
\begin{eqnarray}
P(E,\omega )&=& \sum\limits_{f,i}\int d^3{\bf k}  
|<\Psi^{bulk}_f|\Delta|\Psi^{bulk}_i>|^2
    \nonumber \\ && {}  
\times A_f(E+\hbar\omega) A_i(E)
	         \;
		 \label{A}
		 \end{eqnarray}
which is cast in terms of the one-particle
spectral functions of the initial and final states ($A_i$ and $A_f$), and the 
processes of transport and emission are to be considered separately. 
Assuming that: (i) The system is strictly 2D, and (ii)
the final states form a structureless continuum, 
Eq. 3 reduces to:
$$ P(E,\omega ) \sim  \sum\limits_{i} 
|<\Psi^{bulk}_f|\Delta|\Psi^{bulk}_i>|^2 A_i(E)\eqno(4)$$
in terms of just the $A_i(E)$'s. Further, 
for a single band solid, one obtains:
$$ P(E,\omega ) \sim  
|<\Psi^{bulk}_f|\Delta|\Psi^{bulk}_i>|^2 A_i(E)\eqno(5)$$
where the Fermi function on the right side is suppressed. 

Brief comments on forms 1-5 are appropriate in order to highlight 
the underlying approximations. Since 
forms 3-5 ignore the presence of the surface, it is very difficult 
to include effects of different surface terminations\cite{term}, and the 
associated distortions of bulk wavefunction which can be quite severe. 
In form 4, even near an ARPES peak from a specific initial state, other states 
will in general contribute a background upon being broadened due to 
their finite lifetimes. Forms 4 and 5 ignore the structure in the final state 
spectrum which will be seen below to be quite significant. 
We emphasize that the distinctions between the processes of 
excitation, transport and emission through the surface
invoked by the three-step model (Eqs. 3-5) are artificial
since the more satisfactory one-step formula (Eq. 2) 
does not admit such a decomposition.

The relevant technical details of our 
computations  are as follows. 
In order to keep the problem manageable, the modulation of the lattice 
is neglected 
and the crystal structure of BISCO is assumed to be perfectly 
tetragonal; this still involves 30 atoms per unit cell, and a 
substantial extension of the earlier work on relatively simpler
lattices\cite{stanf,ncco} 
The crystal potential was first obtained self-consistently within
the KKR scheme\cite{kkr}, and essentially yielded the well-known 
LDA-based band structure and Fermi surface 
of Bi2212 \cite{biscoband}; however, the actual potential 
used here is slightly modified in that the Bi-O pockets 
around the $\overline{M}$-point are lifted 
above the Fermi energy to account for their absence 
in the experimental spectra.\cite{foot4}
The surface is assumed to terminate in the Bi-O layer in accord with the 
general consensus\cite{biscocleave} that this is an easy cleavage 
plane in BISCO; the incident light is assumed polarized in the 
$(k_x,k_y)$ plane since the ARPES spectra of interest here 
seem to be insensitive 
to the z-component $A_z$ of the vector potential, and further, 
dielectric effects would in general complicate the relationship between the 
value of $A_z$ inside and outside the crystal\cite{diele}.

The ARPES simulations of Fig. 1 help set the stage for our discussion. 
As expected, the separation between the spectral features A and B 
in Fig. 1a associated with the CuO$_2$ plane bands 
increases as we go from $\overline{\Gamma}$ 
towards $\overline{M}$.\cite{foot45}
Figs. 1b 
and 1c show that the spectral weight of A or B 
depends dramatically on the values of $k_\parallel$
and photon energy $h\nu$. The weights display large changes for 
either peak as a function of $h\nu$ for a fixed $k_\parallel$, or as a 
function of $k_\parallel$ for a fixed $h\nu$. These results 
highlight the importance of matrix element effects 
in BISCO since the weights will be constant 
independent of $h\nu$ or $k_\parallel$ in the absence of these effects. 
In fact, our theoretically predicted $h\nu$ and $k_\parallel$ dependencies 
of spectral weights are in substantial accord with the measurements; 
we discuss this aspect now with reference to Figs. 2-4. 

Fig. 2 compares the measured and computed total spectral 
weights in a 500 meV
energy window below $E_f$ along the three 
high symmetry directions. The characteristic features of the 
experimental data are: a relatively low flat intensity along 
$\overline{\Gamma}\overline{X}$,
a steep rise along 
$\overline{\Gamma}\overline{M}$ compared to $\overline{\Gamma}\overline{Y}$,
and the presence of the prominent bump around $k_\parallel \approx 
0.25 $ along 
$\overline{\Gamma}\overline{Y}$. All these aspects 
of the data are essentially reproduced by the theory. 
\cite{foot47}

Fig. 3 considers the photointensity in the $(k_x, k_y)$ plane for two 
different polarizations of the incident light where the initial 
state is held fixed at $E_f$. The color plots give computed intensities 
over a dense grid of $k_\parallel$-values and display the two CuO$_2$ plane 
band sheets A and B in the band structure; these plots 
are representative of what could be measured in a suitably arranged 
constant-initial-energy angle-scanned (CIE-AS) 
ARPES measurement.\cite{cucis,cieas,foot5} 
Incidently, Figs. 3a2-3c2 are not symmetric about a horizontal 
or vertical line through the center. In Figs. 3b2 and 3c2, the light is 
incident along the $\overline{\Gamma}\overline{Y}$ direction  and the figure
therefore is symmetric only about this diagonal line. In Fig. 3a2, on the 
other hand, the light is polarized horizontally, and the intensities are 
symmetric around the $\overline{\Gamma}\overline{M}$ line; this symmetry 
becomes visible only when the figure is extended in the vertical direction 
to include a larger range of momenta. 

The intensity associated with the outer plane band sheet A as one 
goes around the Fermi surface is shown in quantitative detail in 
Figs. 3a1-3a3. Since A is generally more intense than the 
inner plane band B, A is presumably more relevant in connection with the 
experimental data near $E_f$. For light polarized along the 
horizontal direction (Fig. 3a), we see that the intensity is large 
around $\alpha=0^o$ and decreases rapidly beyond $\alpha \approx 20^o$. 
A 45$^o$ rotation of the polarization vector (Figs. 3b and 3c) induces 
substantial changes in the shape and magnitude of the intensity, and the 
appearance of a minimum at $\beta$ or $\gamma=45^o$. The experimental points 
in Fig. 3 are in good overall accord with the measurements, some 
discrepancies around $\alpha$ and $\beta \approx 0^o$ notwithstanding. 
Thus quite large observed variations (nearly an order of magnitude) in the 
emission intensity from different parts of the Fermi surface are 
mainly the consequence of matrix element effects.

Finally, Fig. 4 considers the photon energy dependence of the emission 
intensity of the spectral feature around the $\overline{M}$-point. 
The theoretical curve displays a prominent peak around 22 eV and 
indicates clearly that the final states in BISCO possess considerable 
structure which is neglected in approximations of Eqs. 5 and 6, 
especially when the matrix element in these equations is further 
replaced by a constant. The experimental points which show the 
presence of a broad peak centered around 21 eV are in good overall 
agreement when we keep in mind that errors of the order of a few 
eV's in locating the {\it final states} are generally expected in the first 
principles band structures. 

We emphasize that the agreement seen in Figs. 2-4 is robust 
to uncertainties inherent in such a comparison on the theoretical 
as well as the experimental side. The experimental weights depend on 
the specific energy window used in their definition, and on whether 
or not a suitable background is subtracted.\cite{foot6}
In this vein, the computed 
weights depend also of course on the specific values of the initial 
and final state damping parameters $\Sigma_i^{''}$ and $\Sigma_f^{''}$. 
In order to assess these effects, we have carried out extensive simulations 
using a variety of different values and energy-dependencies of $\Sigma_i^{''}$ 
and $\Sigma_f^{''}$, in addition to varying the real part of the initial state 
self-energy (in order to mimic correlation effects, even though the 
LDA framework underlies our computations), 
and find that the main features of the results of Figs. 2-4 
are insensitive to such variations.

In summary, we have carried out extensive first-principles one-step
ARPES simulations in Bi2212 wherein the photoemission process is 
modelled realistically by taking into account the nature of 
the initial and final state crystal wavefunctions as well as the 
multiple scattering effects in the presence of a specific 
surface termination. We focus on the nature of the ARPES feature 
arising from $CuO_2$ plane bands, and consider in particular its 
spectral weight as a function of $k_\parallel$ as well as the 
energy and polarization of the incident photons. Large 
variations in the spectral weights predicted theoretically 
along three different high symmetry directions 
are in good accord with the corresponding 
measurements. A good agreement between theory and experiment 
is also seen with regard to changes in spectral weights with 
photon energy around the $\overline{M}$-point, as well as along the Fermi
surface contours in the $(k_x, k_y)$ plane for two different 
polarizations. This study shows 
clearly that the remarkable observed changes in the ARPES spectral 
weights in Bi2212 are essentially a matrix element effect and that 
the importance of matrix elements should be kept in mind in analyzing 
the ARPES spectra in the high-Tc's. 
Another notable implication of this work is that the 
integral (over energy) of the ARPES intensity does not yield the 
momentum density of the electron gas.

This work is supported by the US Department of Energy under
contract W-31-109-ENG-38, including a subcontract to
Northeastern University, and the allocation of supercomputer 
time at the NERSC and Northeastern University Advanced Scientific 
Computation Center (NU-ASCC).

 \end{multicols}
\vskip 2cm
\centerline{\Large \bf Figure captions}
\vskip .5cm
{\noindent {\bf Figure 1:}}
 (a): Simulated ARPES intensity in BISCO for 
 $k_\parallel$ 
 varying between $\overline{\Gamma}$ and $\overline{M}$ (bottom to top) 
 at $h\nu=22 eV$. 
 (b) and (c) show computed variations in spectral 
 weights (areas under peaks) of 
 the two CuO$_2$ plane band features A and B 
 with photon energy for different $k_\parallel$-values, based 
 on a crystal potential without Bi-O hole 
 pockets at $\overline{M}$.
 A small value of the initial and final state damping parameters 
 is used to highlight spectral features. \\

{\bf Figure 2:}
Theoretical weights obtained by integrating the 
$h\nu=22 eV$ ARPES spectra over the binding energy range of 0-500 meV 
are compared with the corresponding experimental 
results \cite{foot1}. 
Curves for the three symmetry lines (Brillouin zone 
in inset) are offset vertically for clarity;
$k_\parallel$ is given in relative units
such that the distance from $\Gamma$ to X, Y or M is defined to be unity
for each direction. Theoretical values are normalized to match 
the experimental value at the maximum around $k_\parallel\approx 0.8$ in the 
$\overline{\Gamma}\overline{M}$ curve.

{\bf Figure 3:}
The color plots give simulated ARPES intensities 
for emission from $E_f$ at $h\nu=22$ eV 
for two different polarizations (white arrows) of the incident 
light. The CuO$_2$ plane band sheets are marked A and B. 
The intensity (area under peak) of the outer plane 
band A is shown in (a1)-(c1) as a function of the 
angles $\alpha$, $\beta$ and $\gamma$. 
Experimental data after Ref. \onlinecite{foot1}.
Theory normalized around $\alpha\approx 0 $ as shown in (a1).

{\bf Figure 4:}
Spectral weight (over a 500 meV window) 
of the feature at $\overline{M}$-point is compared with the 
corresponding experimental\cite{foot1} results as a function 
of the photon energy. Theory normalized to experiment around 
21 eV.


\begin{references}
\bibitem{gen} See, e.g. the collected volumes of Refs. 2-4. 
\bibitem{cape}Spectroscopies in Novel Superconductors, edited by
   A. Bansil, R. Markiewicz, S. Sridhar 
   and D. Liebenberg[J. Phys. Chem. Solids {\bf 59}, No 10-12(1998)].
\bibitem{stand}Spectroscopies in Novel Superconductors, edited by
   Z.-X. Shen, D.H. Liebenberg 
   and A. Bansil[ J. Phys. Chem. Solids {\bf 56}, No 12(1995)].
\bibitem{santa} Spectroscopies in Novel Superconductors, edited by
   F. M. Mueller, A. Bansil and A. J. Arko[J. Phys. Chem. 
   Solids {\bf 54}, No 10(1993)].
\bibitem{Pendry2} J.B. Pendry, Surface Sci. {\bf 57} 679(1976).
   J.F.L. Hopkinson, J.B. Pendry and D.J. Titterington,
   Computer Phys. Commun.  {\bf 19}, 69(1981).
\bibitem{mlcape}A. Bansil and M. Lindroos, 
   J. Phys. Chem. Solids {\bf 59}, 1879(1998).
\bibitem{ncco}M. Lindroos and A. Bansil, Phys. Rev. Lett. {\bf 75}, 1182(1995).
\bibitem{cucis} M. Lindroos and A. Bansil, 
   Phys. Rev. Lett. {\bf 77}, 2985(1996).
\bibitem{prl} K. Gofron, J. C. Campuzano, A. A. Abrikosov, M. Lindroos,
   A. Bansil, H. Ding, and B. Dabrowski, Phys. Rev. Lett. {\bf 73}, 3302(1994).
\bibitem{stanf}A. Bansil and M. Lindroos, J. Phys. Chem. Solids
     {\bf 56}, 1855(1995).
\bibitem{foot1} J. C. Campuzano and H. Ding, private communication; data 
   based in part on results of Ding et al.[Phys. Rev. Lett. 76, 1533(1996)]. 
\bibitem{foot2} As an operational definition, by ``matrix element effects"
   we mean the effect of approximating the true spectral photointensity
   of say Eq. (2) by the bulk initial state spectral density function $A_i(E)$. 
\bibitem{anderson1} P.W. Anderson and Y. Ren in, High temperature 
   superconductivity, eds. K. S. Bedell et al.
   (Addison-Wesley, Redwood City, 1990). 
\bibitem{Schaich} N.W. Ashcroft, W.I. Schaich, Phys.Rev {\bf B3},2452(1971).
\bibitem{Kunz} C. Kunz in, Photoemission in Solids II, Eds. L. Ley 
   and M. Cardona, Springer-Verlag, Berlin, 1979, p. 313.
\bibitem{term}M. Lindroos et al., Physica {\bf C212}, 347(1993).
\bibitem{kkr}
   A. Bansil, S. Kaprzyk, and J. Tobola, MRS Proc. {\bf 253}, 505(1992);
   A. Bansil and S. Kaprzyk, Phys. Rev. {\bf B 43}, 10335(1991);
   S. Kaprzyk and A. Bansil, Phys. Rev. {\bf B 42}, 7358(1990).
\bibitem{biscoband}
   S. Massida, J.Yu, and A.J. Freeman, Physica {\bf C 152},251(1988); 
   H. Krakauer and W. E. Pickett, Phys. Rev. Lett. {\bf 60},1665(1988).
   M. S. Hybertsen and L.F. Mattheiss, Phys. Rev. Lett. {\bf 60},1661(1988).
\bibitem{foot4} Although the ARPES data from BISCO 
   suggest the absence of Bi-O pockets around $\overline{M}$, 
   the conclusions of this article are insensitive to whether or not this
   is so. 
\bibitem{biscocleave}See, e.g. 
   D.M. Ori, A. Goldoni, U. del Pennino and F. Parmigiani, 
   Phys. Rev.  {\bf B 52}, 3727(1995).
\bibitem{diele} See, e.g. A. Gerlach, R. Matzdorf and A. Goldmann, 
   Phys. Rev.  {\bf B58},10969(1998).
\bibitem{foot45}Experimental spectra do not obviously indicate
the presence of a feature
like C or C' close to $E_f$. However, the location of C, C' is 
quite sensitive to the crystal potential in the Bi-O planes which 
has been modified somewhat in these simulations to move the 
Bi-O pockets above $E_f$.
\bibitem{foot47} The discrepancy in Fig. 2 between 0.4-0.8 along
$\Gamma M$ is related to the structures C and C' in theory and is not 
considered serious; see also Ref. 22 above. 
\bibitem{cieas}
   N. L. Saini et al., Phys. Rev. Letters {\bf 79}, 3467(1997). 
\bibitem{foot5} The finer details (as in the data of 
   Ref. \onlinecite {cieas}) associated with 
   modulations and possible stripe phases are of course not reproduced 
   in the color plots of Fig. 3 which assume a perfect tetragonal lattice. 
\bibitem{foot6}Although experimental data after Ref. 11 is used in Figs. 2-4, 
results from various groups are similar, see, e.g. 
Loeser et al.\cite{loeser}; Chuang et al. \cite{Chuang}; Saini et al.[24]. 
\bibitem{loeser}A. G. Loeser et al., Phys. Rev. {\bf B 56}, 14185(1997).
\bibitem{Chuang}Y. -D. Chuang  et al. (preprint cond-mat/9904050).
 \end{references}
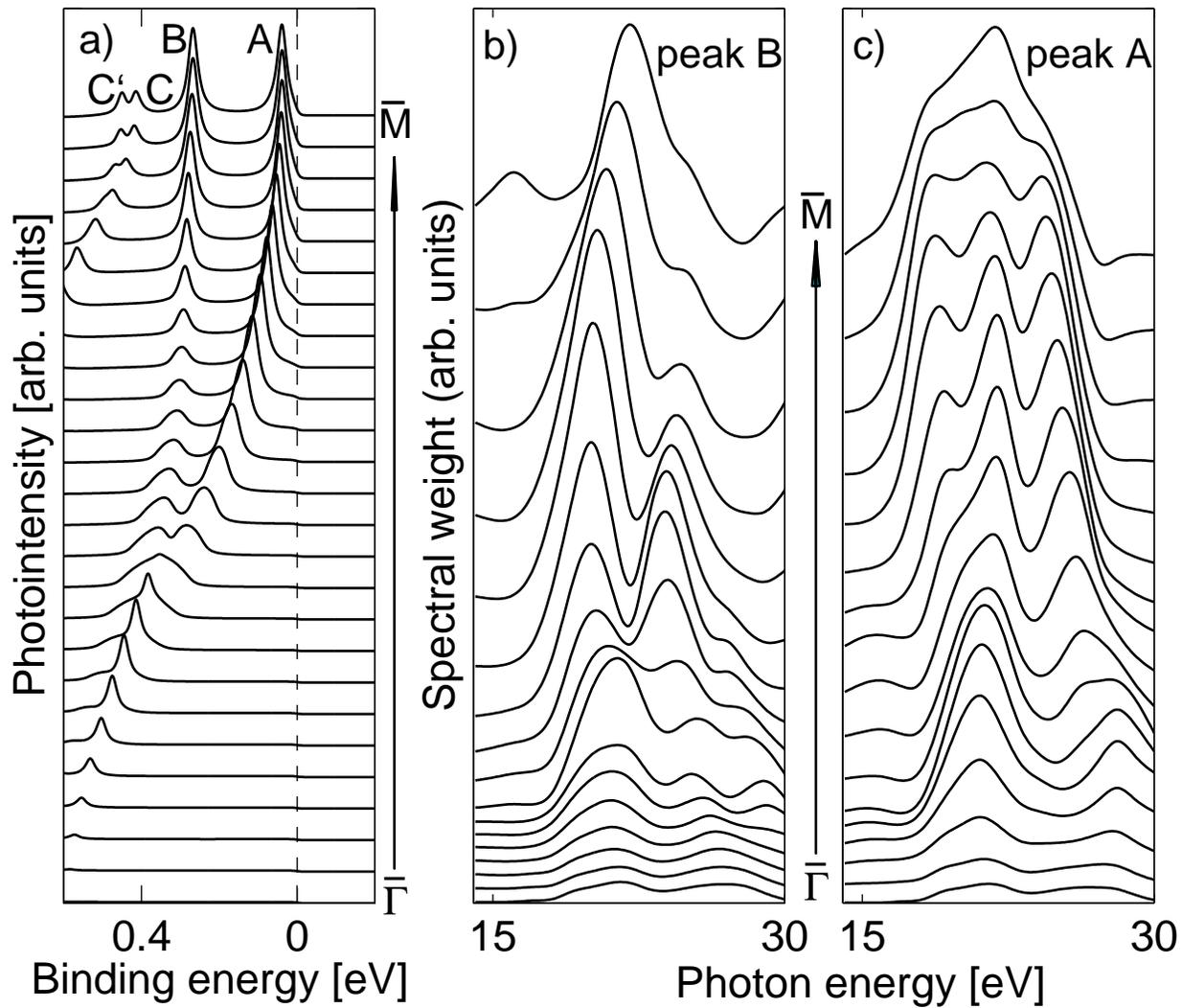
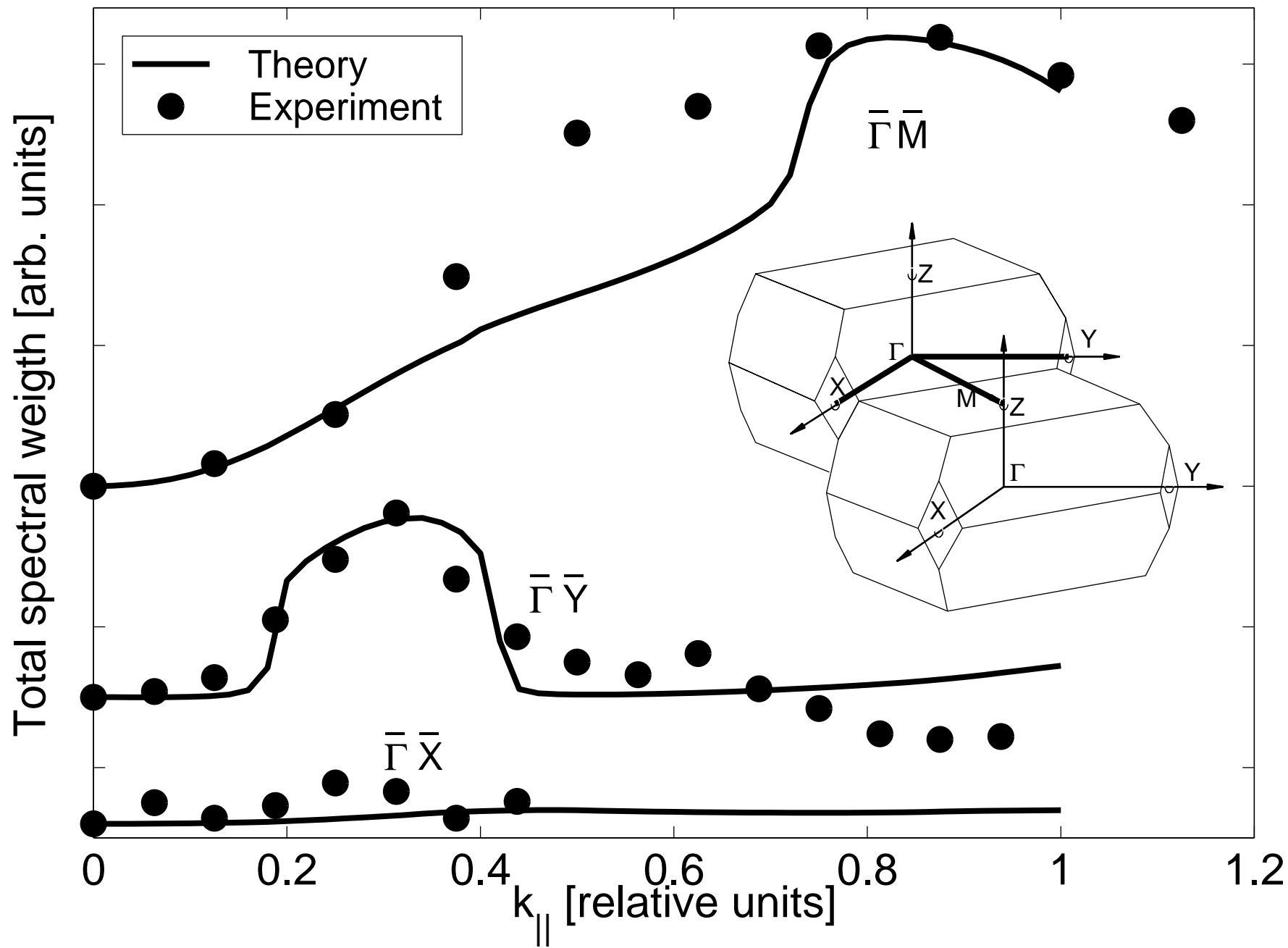
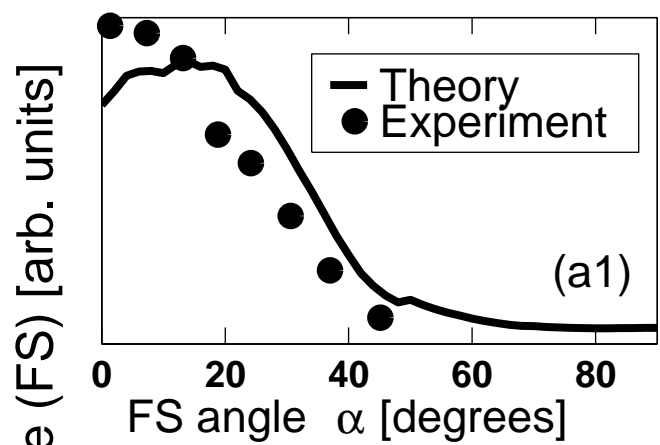
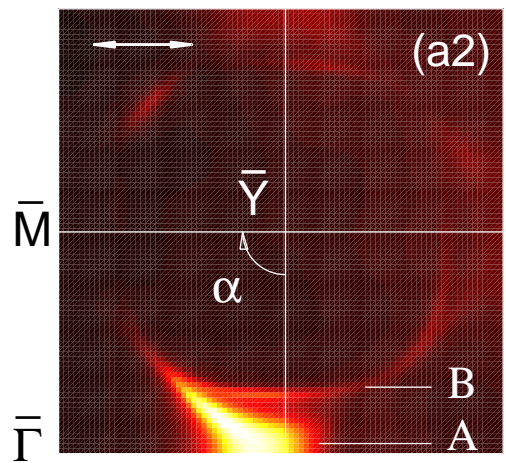
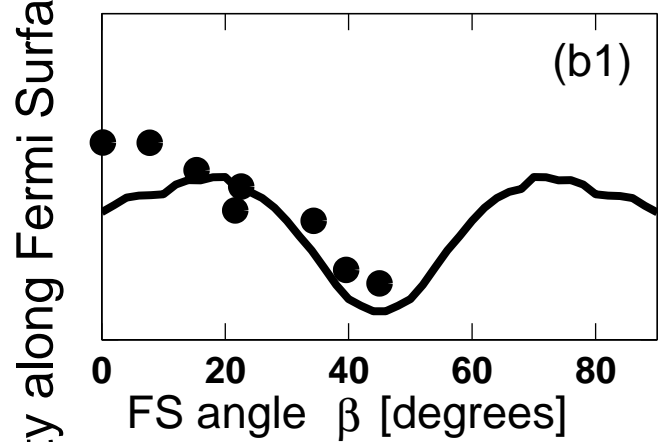
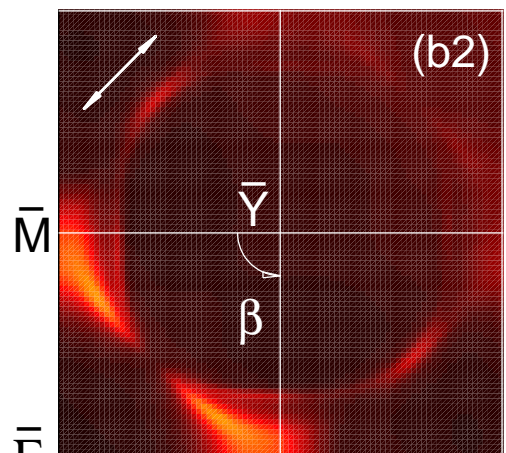
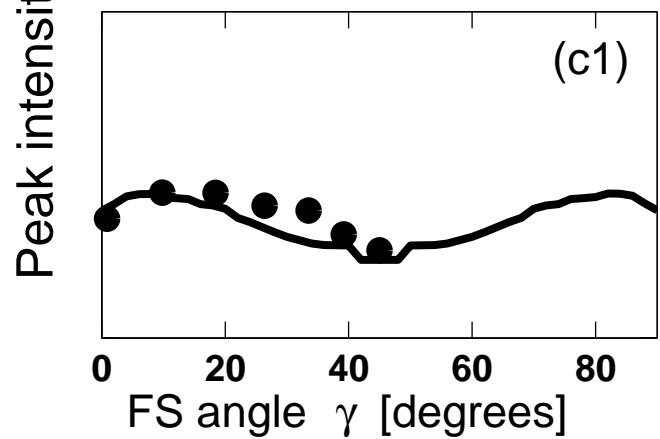
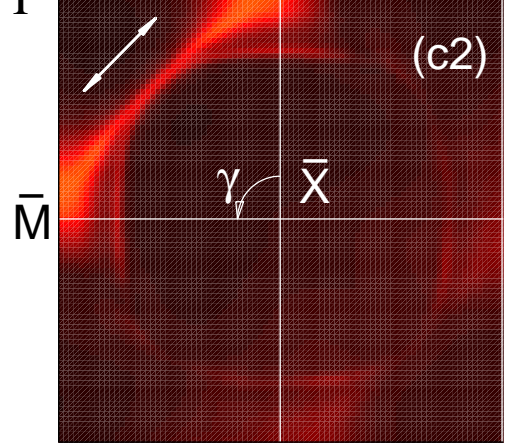
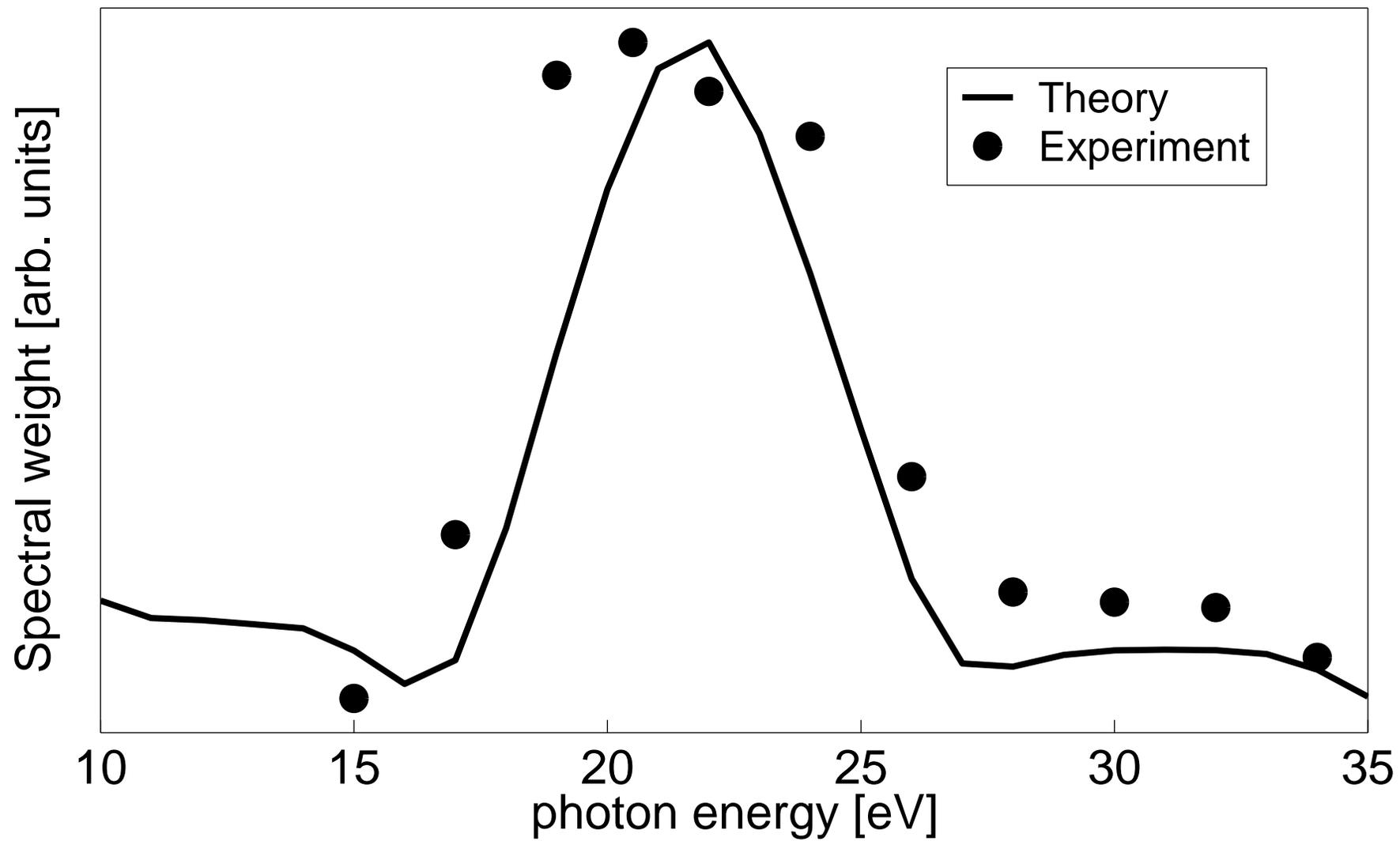
\end{document}